\documentstyle[prl,aps,psfig,multicol]{revtex}  

\newcommand{\be}{\begin{equation}}
\newcommand{\ee}{\end{equation}}
\newcommand{\bes}{\begin{eqnarray}}
\newcommand{\ees}{\end{eqnarray}}

\begin{document}

\title{The Bak-Chen-Tang Forest Fire Model Revisited}
\author{Hans-Martin Br\"oker$^1$ and Peter Grassberger$^{1,2}$}
\address{ 
$^1$ Physics Department, University of Wuppertal, D-42097 Wuppertal, Germany\\
$^2$ HLRZ c/o Forschungszentrum J\"ulich, D-52425 J\"ulich, Germany}
 
\date{\today}
\maketitle

\begin{abstract}

We reconsider a model introduced by Bak, Chen, and Tang (Phys. Rev. {\bf A 
38}, 364 (1988)) as a supposedly
self-organized critical model for forest fires. We verify again that
the model is not critical in 2 dimensions, as found also by previous 
authors. But we find that the model does show anomalous scaling 
(i.e., is critical in the sense of statistical mechanics) in 3 and 4 
dimensions. We relate these results to recent claims by A. Johansen.
 
\end{abstract}
 
\begin{multicols}{2}

During the last ten years, the concept of self-organized criticality (SOC),
proposed by Bak, Tang and Wiesenfeld in \cite{bak1}, has been applied to 
a large number of phenomena. Although a generally accepted and rigorous 
definition of SOC still 
doesn't exist, a number of general features are supposed to hold for any 
model which shows SOC (see also \cite{flyv}): \\
(i) There should be scaling laws which in general will be anomalous for 
local models (for mean field or random neighbor models the exponents in 
general will be integer or half-integer, i.e. `normal'). These scaling 
laws should not be `trivial' (such as $m \sim l^3$ for the mass-length 
connection in 3-d), though triviality is not always easy to define. \\
(ii) There should be no control parameter which has to be fine tuned. 
Thus the scaling should be a robust phenomenon, in contrast to standard 
critical phenomena. Again this criterion is less clear-cut then one 
might wish. In particular, \\
(iii) SOC shows up usually in slowly driven systems, when the driving 
rate tends to zero (an advocatus diaboli who insists that this rate should 
be considered as a control parameter could thus conclude that SOC doesn't 
exist at all). Typically, such systems become locally unstable when the 
stress exerted by the driving exceeds some limit, and react with `avalanches'
of activity which are large on microscopic scales, but small on macroscopic 
ones. \\
(iv) Finally, a common feature of many systems with SOC is that the driving 
is not controlled as usual through a {\it force}, but through a {\it flux} 
(i.e., an extensive quantity) \cite{sornette}.

Phenomena where these features show up include sand piles \cite{bak1}, 
earth quakes \cite{cl,ocf}, pinned surfaces \cite{buldy,sneppen},  and
biological evolution \cite{bs}. A last application are forest fires with 
small growth rate of trees and even smaller rate for spontaneous ignition 
(`lightning') \cite{ds}. The latter example is special in the sense that 
it requires {\it three} different time scales for criticality.

In \cite{bak2} (BCT), it was claimed that this list should also include 
forest fires without lightning. The specific model studied by BCT used 
a regular $d$-dimensional lattice and discrete time, with each lattice site
in one of 3 possible states: green tree, burning tree, or ash. During one 
time unit, a burning tree ignites all green neighbors (if there are any) 
and turns itself into ash. This is the fast part of the dynamics. The slow 
part is the re-growth of trees. It is modelled by a stochastic spontaneous 
transition ash $\to$ tree with probability $p\ll 1$. Thus in each time step 
a randomly selected fraction $p$ of all ash sites is flipped into trees.
As pointed out in \cite{gk}, a much better 
interpretation of the BCT model would be in terms of epidemics 
with slow recovery resp.~slow loss of immunization, but we shall continue 
to speak about forest fires for convenience. 

Simulations \cite{gk} on large lattices (up to $4800^2$ sites) and for very 
small values of $p$ (down to $5\times 10^{-4}$) showed however that dynamics 
in $d=2$ is quite different from that suggested by BCT on the basis of  
small-scale simulations. Instead of being critical, 
the system develops noisy spiral patterns which become less and less noisy 
as $p\to 0$. More precisely, spiral arms (fire fronts) propagate with 
finite mean velocity $v$
for any $p$, and the typical distance $\xi$ between spiral arms (the 
characteristic length scale) as well as the time $T$ between two passings of 
a front scale as $1/p$: $\xi\propto T \propto 1/p$. In the limit $p\to 0$ 
the coherence length thus diverges, implying that the dynamics is governed 
by {\it average} tree densities over larger and larger regions. In this 
limit the dynamics is thus similar to that of coupled relaxation oscillators 
with very sudden discharge and very slow recharging. This explains 
qualitatively the patterns found, though any details are still badly 
understood due to the inherent noise for any $p\neq 0$ which leads to 
permanently ongoing pattern rearrangements. 

Indeed, the authors of \cite{gk} were rather careful in the interpretation 
of their simulations. They pointed out that {\it if} the model were 
non-trivially critical with characteristic length and time scales scaling 
as 
\be
   \xi\propto p^{-\alpha},\;\; T\propto p^{-\beta}\qquad \alpha,\beta \neq 1,
\ee
then this could only be the case if fire fronts get fuzzy for $p\to 0$, and 
front velocities would tend to zero. In such a case the fire would be at 
a critical 
{\it percolation} threshold (a similar scenario does hold indeed for some 
versions of forest fire models with lightning \cite{drossel,css}). The medium 
into which it percolates would not be uncorrelated, in contrast to standard 
percolation. Using nevertheless estimates of critical exponents from the 
latter, it was concluded in \cite{gk} that such an alternative scenario 
was very unlikely but hard to exclude rigorously. The same conclusion (with 
less caveats in spite of less statistics) was reached also in \cite{mds}. 

Simulations in 3 and 4 dimensions were much 
less significant. This was partly due to difficulties in visualizing 
such systems, partly because it is hard to keep a fire from getting 
extinct for small values of $p$. Nevertheless some indications for 
well defined fire fronts were seen in $d=3$ \cite{gk}, and it was concluded 
that basically the same scenario holds there as in $d=2$. Obviously, this 
cannot be true for arbitrarily high dimension. At least for $d\geq 6$ the 
behavior should be that of dynamic percolation, with small fluctuations 
of the density of trees around a mean field value. The characteristic 
time of these oscillations is not $T \sim 1/p$ as for fronts, but 
$T \sim 1/\sqrt{p}$ \cite{gk}. Simulations showed the actual behavior 
for $d=3$ and $d=4$ was in between both \cite{gk,mds}. Unfortunately, this 
was not followed up, and the possibility $T \sim 1/p^\beta$ with $1/2 < 
\beta < 1$ was not considered seriously.

In a recent paper, Johansen \cite{aj} claimed exactly that. In addition, 
he claimed that the same is true also in $d=2$. On the other hand he 
confirmed that spirals are formed in $d=2$, and that the typical distance 
between fire fronts scales as $L\sim 1/p$. Now it is easy to see that the 
latter statements are self-contradictory. They would imply that fronts 
propagate with speed $v = L/T \sim p^{-(1-\beta)} \to \infty$ for $p\to 0$.
Since the front can propagate at most one lattice site in each time 
step, this is impossible.

In order to clarify this situation, we report in the present letter on 
simulations where we measured $T$ with high precision, for 
$d=2,3$ and 4, and for wide ranges of $p$. For $d=3$ we find indeed a 
very clear indication of anomalous scaling, $\beta = 0.77\pm 0.02$. The 
situation is slightly less clear in $d=4$ for reasons detailed below, but 
we again find scaling (with $\beta \approx 0.6\pm 0.05$). These findings 
imply that $pT \to 0$ for $p\to 0$. For $d=2$, finally, we find that $pT$ 
also decreases slightly with decreasing $p$, but not as a power law. 
Our data are not precise enough to distinguish clearly 
between a logarithmic increase, 
\be
   pT \sim [1/\log(1/p)]^\gamma,\quad \gamma>0 ,     \label{pT}
\ee
and a limited increase which leads to a finite value at $p\to 0$. They 
favor the latter. But if eq.(\ref{pT}) would hold, instead, we would 
have the alternative scenario mentioned above in which fronts become 
fuzzy, front velocities become zero, and the evolution is basically 
a critical (correlated) percolation phenomenon. 

The latter seems to apply in $\geq 3$ dimensions, although 
the situation is not clear there either. The problem is the following: 
if $pT \to 0$ for $p\to 0$, then the fraction of replenished trees between 
two peaks of the local fire intensity has to go to zero also. Except 
for transients this means that also the fraction of trees burnt during 
such a peak must tend to zero. In such a case we would naively expect 
that the amplitude of any (noisily) periodic observable should diminish 
when $p$ is decreased. This is not observed. Instead, the peaks stand 
out {\it very} clearly, even for the smallest values of $p$. Thus 
our data suggest 
at first sight that fires burn large areas completely, also in 3 and 4 
dimensions (they do so in $d=2$, but there it is expected). This 
might indicate that even with $p$ values as small as $10^{-3}$ 
we are not yet in the asymptotic scaling region, 
but an alternative scenario will be discussed below. 

\smallskip

For our simulations we used the basic algorithm described in 
\cite{gk}. The current state of the system is encoded in two data 
structures: a list of burning tree sites, and a bit map indicating for 
each site whether it contains a tree or not. Notice that we do not 
have to distinguish in the latter between burning and non-burning trees 
since that information is contained in the list. Thus we can use bit 
coding in order to simulate very large lattices. Instead of using $d$ 
coordinates, each site is indexed 
by a single integer, and boundary conditions are helical. In each 
time step, a new list of burning trees is established by scanning 
through all neighbors of all entries in the old list, and the old 
list is thereafter replaced by the new one. After this, $pN_{\rm ash}$ 
sites ($N_{\rm ash}$ is the total number of sites containing ash) are 
randomly selected and switched from ash to tree. 

To avoid that the fire dies out, we used correlated 
random initial conditions, and discarded transients which involved 
at least 100 oscillation periods. If the fire died out nevertheless, 
we ignited some fires `by hand', and started the run again.

Lattice sizes went up to $16384\times 16384$ in 2 dimensions (for 
$p=0.00005$), and to comparable numbers of sites in $d\geq 3$. Simulation 
times went up to $t=4\times 10^6$, and were always larger that $100/p$.

Since the aim of the present paper was to obtain precise estimates 
of $T$, the biggest effort was devoted to that. As in previous
analyses, we used the number of burning trees as observable. But in 
order to improve the signal to noise ratio, we did {\it not} simply 
measure the total number, averaged over the entire lattice. The reason 
is that fires in different parts of a large lattice will in general not 
oscillate in phase, whence cross correlations from distant regions 
will mainly contribute to the noise in the autocorrelation function. 
We therefore proceeded as 
follows: we divided the lattice into hypercubes of linear size 
$l$ with $l<L$, and measured the numbers $n_i(t)$ of burning trees 
in the $i$th cube at time $t$. From each of these local time series we 
estimated autocovariances 
\be
   c_i(t) = \langle n_i(\tau) n_i(t+\tau) \rangle
\ee
which were then averaged over the lattice, 
\be
   c(t) = \sum_i c_i(t) \;.
\ee
Oscillation periods $T$ were either estimated from peak-to-peak 
distances in $c(t)$ or by Fourier transforming $c(t)$, obtaining 
thereby maximum entropy spectrum estimate $S(f)$ \cite{press}. 
Results were the same. Two typical plots of $c(t)$ are shown 
in fig.~1.

\begin{figure}[ht]
\psfig{file=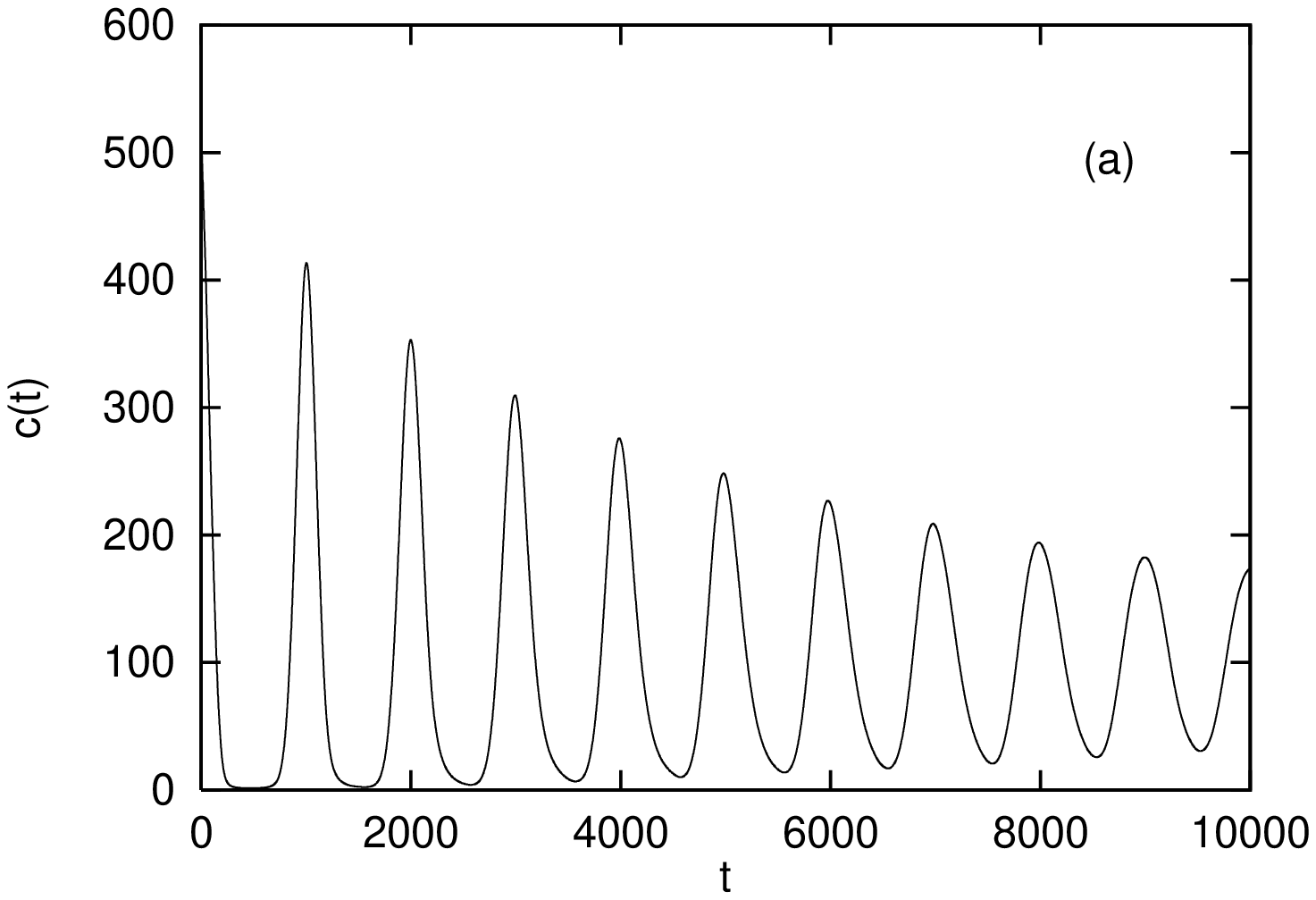,width=6.2cm,angle=270} 
\psfig{file=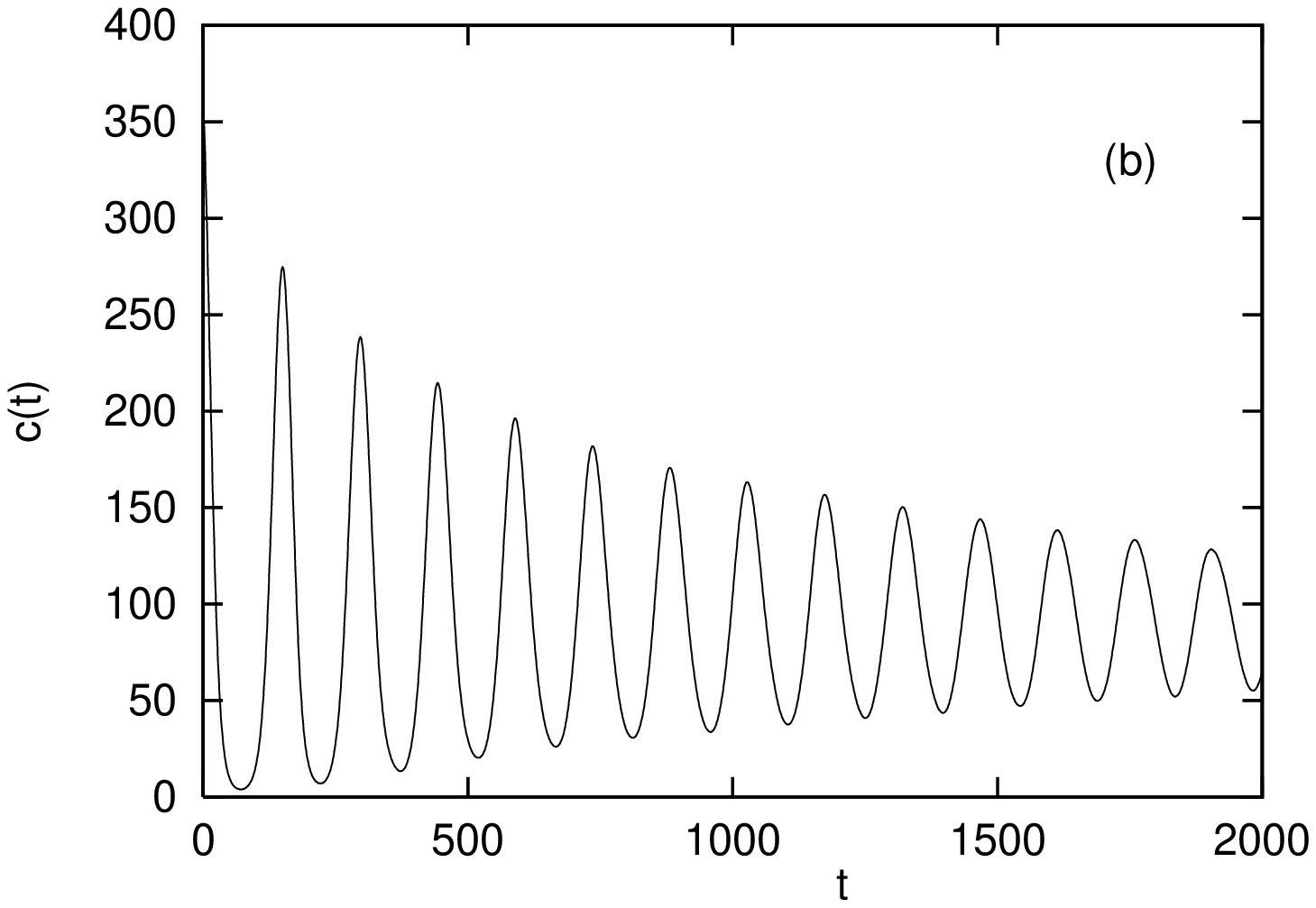,width=6.2cm,angle=270}
{\small FIG. 1: Autocovariances $c(t)$ with: (a) $d=2$, $p = 0.001$, 
lattice size $4096\times 4096$, $10^6$ iterations, and $l = 128$; 
(b) $d=3$, $p=0.003$, lattice size $256^3$, $10^5$ iterations, and 
$l = 16$. In both cases,  only 16 squares resp.~cubes were used in 
the averaging. Normalization is arbitrary.} 
 \label{corr.fig}
\end{figure}

The sharpness of the peaks in $c(t)$ and the strong higher 
harmonics result from the fact that $l<<\xi$, whence $n_i(t)$ is 
non-zero only during the short time when a fire front passes through 
cube $i$. As a consequence we obtain very clean signals and very 
precise estimates of $T$. 

\begin{figure}[ht]
\centerline{\psfig{file=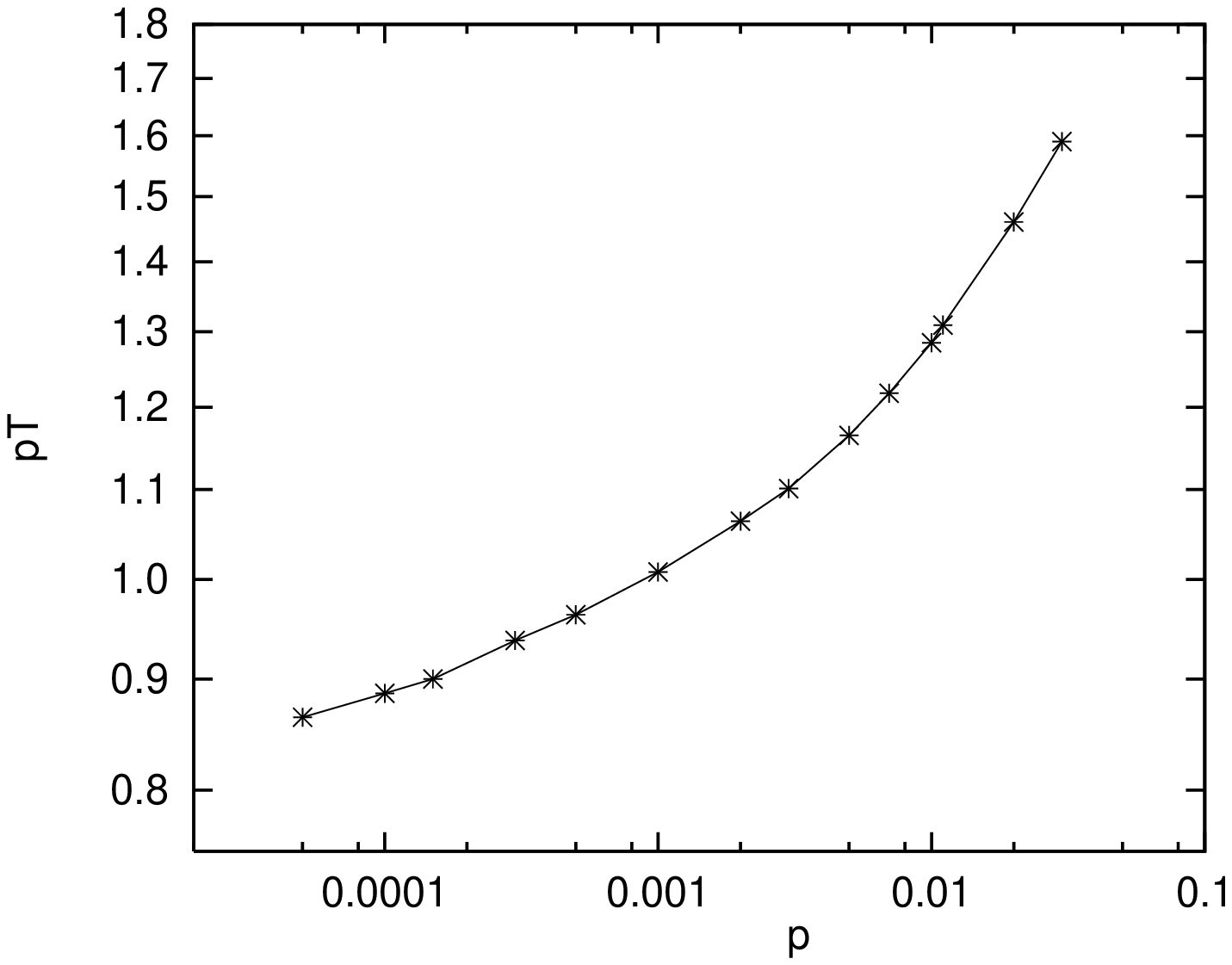,width=7.1cm,angle=270}}
{\small FIG. 2: Log-log plot of $pT$ versus $p$ for 2 dimensions. Here, $T$ 
is the average peak-to-peak distance in $c(t)$.} 
 \label{pt_2d.fig}
\end{figure}

We should point out here that $T$ is not the only time scale characterizing 
$c(t)$. First of all, there is also the coherence time $\tau_{\rm coh}$. 
It can be measured from the asymptotic exponential decay of the 
oscillation amplitudes. As expected, it grows quickly with $1/p$. But 
we did not make systematic measurements since the exponential decay 
is not observed at finite times due to a third time scale, namely the 
regeneration time $\tau_{\rm regen} =1/p$. It is easy to see that the 
amplitudes of all peaks at $t>0$ have to decrease for fixed $l$ and 
$p\to 0$, if $T\ll\tau_{\rm regen}$ in this limit. Indeed, for $l=1$ 
one finds $c(T)/c(0) \approx pT$. Thus the shape of $c(t)$ depends 
strongly on the box size $l$, but we verified that the locations of the 
maxima (which determine $T$) are independent of it.

For small values of $p$ (requiring 
large systems and long simulation times) it was not feasible to store 
all $n_i(t)$ for each cube $i$ and every $t$. In such cases we 
decimated the data by either reading out only a fraction of cubes, 
or by coarse-graining in $t$ and storing $n'_i(kt) = \sum_{\tau=kt}^{
(k+1)t} n_i(\tau)$ for some integer $k>1$. Both gave nearly the same
performance as without decimation, even if the data were reduced 
by more than one order of magnitude.

\begin{figure}[ht]
\centerline{\psfig{file=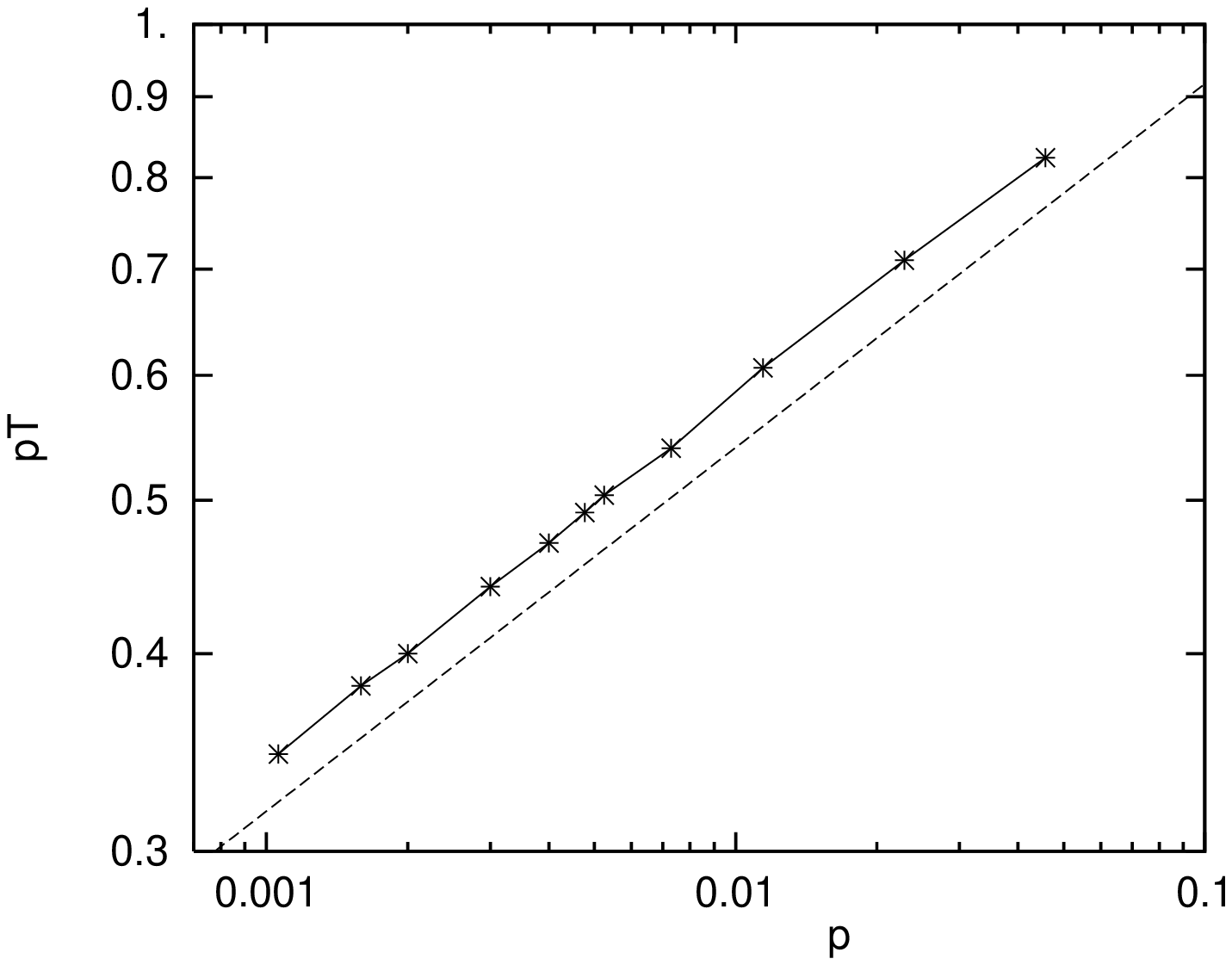,width=7.1cm,angle=270}}
{\small FIG. 3: Same as fig.~2, but for \protect{$d=3$}. 
The dashed line corresponds to a power law \protect{$T \sim p^{-\beta}$} 
with \protect{$\beta = 0.77$}.} 
 \label{pt_3d.fig}
\end{figure}

Our final results are shown in figs.~2 to 4. Each of 
them is a log-log 
plot showing $pT$ versus $p$. In figures 2 and 3 the 
estimated errors are smaller than the symbols. We see clearly the 
trends discussed above. For $d=4$ (fig.~4) the errors are larger 
and not purely statistical: runs with different initial conditions 
gave occasionally values which differed by more than naive 
statistical error estimates. Obviously this means that either the 
system is not ergodic, or that our simulation times were not 
sufficient to explore all phase space. Nevertheless we believe that 
the data give a clear indication of scaling also for $d=4$.

\begin{figure}[ht]
\centerline{\psfig{file=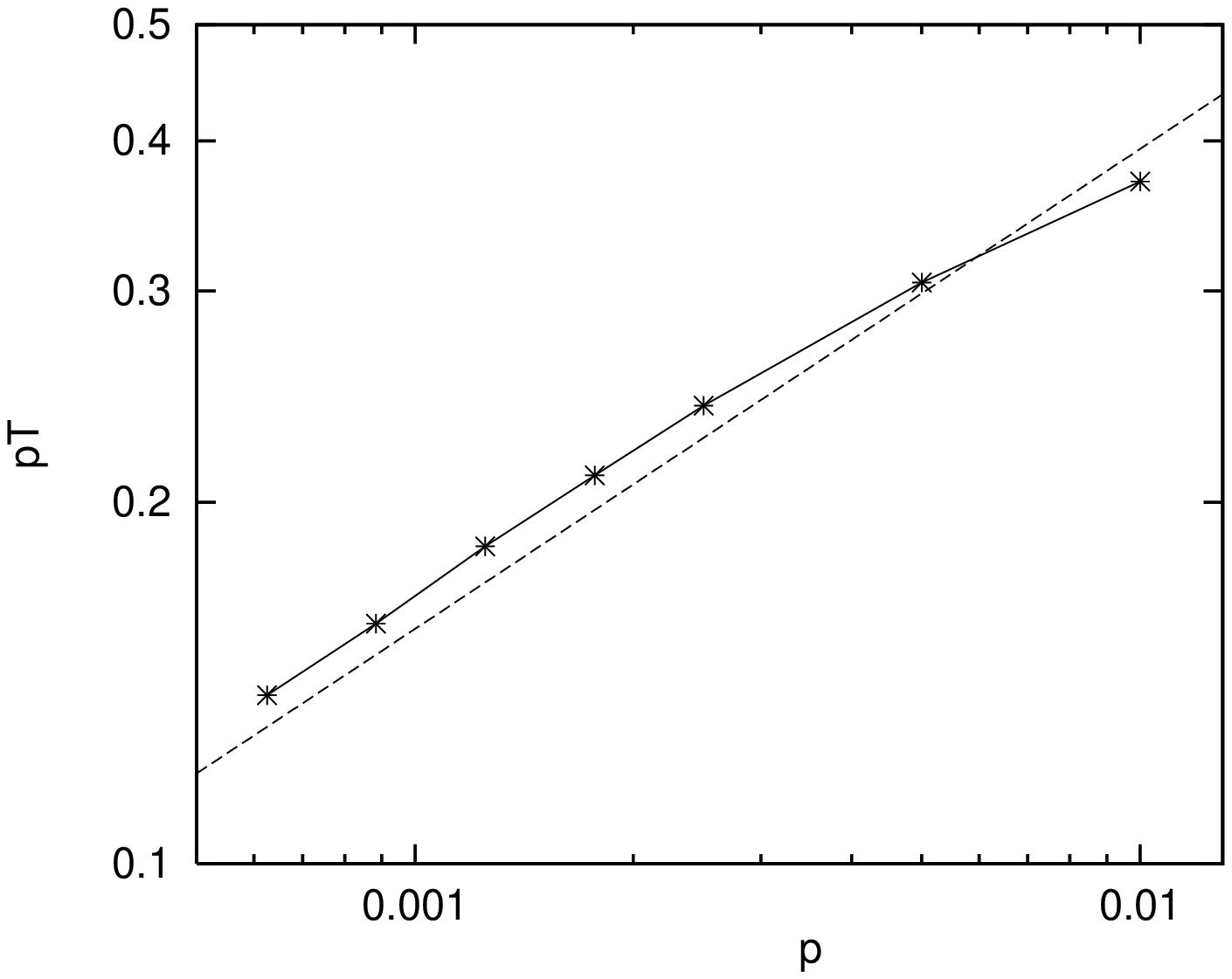,width=7.1cm,angle=270}}
{\small FIG. 4: Same as fig.~2, but for \protect{$d=4$}. 
The dashed line corresponds to \protect{$\beta = 0.6$}.} 
 \label{pt_4d.fig}
\end{figure}

As pointed out above, the most straightforward interpretation of our 
data for $d\geq 3$ is that there are fronts with distances $\xi\approx 
vT \sim p^{-\alpha}$. Between two successive fronts only a tiny fraction 
of trees can re-grow, whence also a tiny fraction of sites could 
burn when the front passes. If these fronts would propagate into an 
essentially unstructured medium, this would imply that the process is 
close to critical percolation, and it would be hard to understand the 
very large and regular amplitudes seen e.g. in fig.1b. To understand 
better what is going on we tried to visualize typical 3-d configurations, 
but only with moderate success. But the above suggests that fronts do 
not propagate into an unstructured medium. It is well known that the 
complement of a slightly supercritical percolation cluster in 
$d\geq 3$ is connected. Thus a passing fire, even if it is 
supercritical and has thus a sharp front, could still leave intact 
connected regions with slightly subcritical or even supercritical tree 
densitites. This would allow the next front to pass very soon again. 
The crucial point is that this front should propagate only on a sparse 
but connected subset of trees, requiring the patters of trees to be 
highly structured.

In summary, we have shown that there are anomalous scaling laws in 
the Bak-Chen-Tang forest fire models, but only in higher dimensions. 
This suggests that there are sharp fire fronts in $d>2$, even in 
the limit $p\to 0$, but each front propagates only on a sparse 
subset of trees. In this way the fire can be endemic in $d\geq 3$,
burning only an infinitesimal fraction of trees between two 
recurrences to the same mesoscopic region without, nevertheless, 
resembling critical percolation. 

Since the present model can be considered as a model for extremely 
noisy coupled relaxation oscillators, it is an interesting question 
whether this is the generic behavior of noisy coupled relaxation 
oscillators in $>2$ dimensions.

Another final remark concerns the relationship of the present 
model with other SOC models. The main difference is that in the 
present model the system is slowly driven (by the growth of trees) 
into a {\it susceptible} state, while most other SOC models are 
driven into {\it unstable} states which will `topple' (discharge, 
catch fire, ...) spontaneously. This difference would be blurred 
if a susceptible site has a small chance to topple anyhow, as in 
the Drossel-Schwabl \cite{ds} model. But it is also blurred if the 
connectivity of the lattice is so high that activity can efficiently 
spread over large distances without leaving many traces. This is 
obviously what happens in the present model when $d\geq 3$. This 
explains why the model shows SOC for $d\geq 3$, but not for $d=2$.

\bigskip

Acknowledgment: 
We are indebted to H. Flyvbjerg for useful discussions and a critical 
reading of the manuscript.
This work was supported by the DFG within the Graduiertenkolleg
``Feldtheoretische und numerische Methoden in der Elementarteilchen- und
Statistischen Physik'', and within Sonderforschungsbereich 237.

\end{multicols}
\end{document}